%% file: ms_basic.tex
\documentclass[preprint]{aastex631}
\usepackage{showyourwork}
\usepackage{lineno}

\begin{document}
	
\newcommand{\ergsec}{erg s$^{-1}$ cm$^{-2}$ $\mu$m$^{-1}$}

% Title
\title{Keck Near-Infrared Detections of Mab and Perdita}

% Author list
%\author{@emolter}
\author[0000-0003-3799-9033]{Edward M. Molter}
\affiliation{Earth and Planetary Science Department, University of California, Berkeley, CA 94720, USA}
\correspondingauthor{Edward M. Molter}
\email{emolter@berkeley.edu}

\author[0000-0002-4278-3168]{Imke de Pater}
\affiliation{Earth and Planetary Science Department, University of California, Berkeley, CA 94720, USA}

\author[0000-0002-6293-1797]{Chris Moeckel}
\affiliation{Earth and Planetary Science Department, University of California, Berkeley, CA 94720, USA}

%\linenumbers

\begin{abstract}
We report the first near-infrared detection of Uranus's tiny moon Mab, the presumed source of the blue and diffuse $\mu$ ring, using the NIRC2 instrument at Keck Observatory. The detection was permitted by an updated shift-and-stack procedure allowing us to integrate on Mab as it moved across the detector in 23 separate exposures taken over $\sim$2 hours, as well as the very low (0.02$^\circ$) phase angle at the time of observation.
At this phase angle, Mab has an integrated I/F of 
\input{output/Mab_urh_intif.txt} $\pm$ \input{output/Mab_urh_intiferr.txt} km$^{2}$ 
at 1.6 $\mu$m and 
$\lesssim$\input{output/Mab_urk_intif.txt} km$^{2}$ 
at 2.1 $\mu$m. 
Comparing these values with Mab's visible reflectance as derived by HST reveals that Mab is spectrally blue; its (0.5 $\mu$m)/(1.6 $\mu$m) color is more consistent with Miranda's value than Puck's value. Mab is therefore more likely a $\sim$6-km radius body with a Miranda-like surface than a 12-km radius body with a Puck-like surface, in agreement with prior work based on infrared upper limits, but we caution that a Puck-like color is only ruled out at the 2$\sigma$ level.
We also report the first infrared photometry of Perdita, finding an integrated I/F of 
\input{output/Perdita_urh_intif.txt} $\pm$ \input{output/Perdita_urh_intiferr.txt} km$^{2}$ at 1.6 $\mu$m.
\end{abstract}


% Main body with filler text
\section{Introduction}
\label{s:intro}

% does the below text fit in anywhere?
The small regular moons comprising the Portia group in the Uranian system represent the most densely-packed set of satellites known, with semimajor axis separations of 3\% on average \citep{showalter20}; as such, the system is predicted to be chaotic, with collisions occurring on 10$^5$- to 10$^7$-year timescales \citep{duncan97, french12}. Indeed, the faint dusty $\nu$ ring, situated between the orbits of Portia and Rosalind, has been proposed to be the remnant of such a collision \citep{showalter20}. Due in part to this small orbital spacing, the Uranian satellites likely affect each other's surface properties via dust transport in the circumplanetary environment. For example, dust deposition from Uranus's outer irregular satellites has been invoked to explain spectral reddening of the leading hemispheres of Ariel, Umbriel, Titania, and Oberon \citep{cartwright18}. Material transport is important for understanding the surfaces of Miranda, Puck, and the smaller inner moons, too, but these processes are less well-studied. 
Puck's leading hemisphere has a significantly bluer spectral slope than its trailing hemisphere, meaning that it may be sweeping up material from the $\mu$ ring \citep{french17}. %this is a DPS abstract - is there a better reference?
Unusually low depth/diameter ratios of craters on the potential ocean world Miranda suggest the deposition of a $\sim$1 km thick regolith \citep{cartwright21, beddingfield22}. Those authors propose three possible sources of the regolith material: deposition from the $\mu$ ring, plume deposits from a subsurface ocean, and/or giant impact ejecta.

The faint, diffuse $\mu$ ring is therefore crucial for understanding the evolution of Uranus's circumplanetary system as a whole, but its properties and origin remain poorly understood.
It is the only ring in the solar system with a blue spectrum other than Saturn's E ring \citep{depater06}, which is sourced from Enceladus's cryovolcanism \citep{pang84, porco06, spencer06, spahn06}. The E~ring is blue due to its abundance of tiny ($\lesssim$1 $\mu$m) particles and paucity of larger grains \citep{showalter91, depater04, hillier07}. The tiny moon Mab is embedded in, and is the presumed source of, the $\mu$ ring, but active cryovolcanism on such a small body seems highly unlikely. Instead, interplanetary dust and/or charged plasma impacting Mab's surface may liberate enough particles to form the $\mu$ ring, as suggested by \citet{depater06} based on a model that was originally proposed to explain the blue color of the E~ring before Enceladus's geysers were discovered \citep{horanyi92, hamilton94}. This scenario relies on theoretical expectations that $\sim$1 $\mu$m particles are more likely than larger particles to disperse into a ring (instead of re-accreting to the moon) under the combined influences of gravity, radiation pressure, and electromagnetic forces.  Numerical simulations of $\mu$ ring particles sourced by meteoritic collisions with Mab's surface found that $\sim$1 $\mu$m particles are more susceptible than larger particles to eccentricity perturbations and therefore spread out into a diffuse ring, whereas the larger particles remain confined near Mab's orbital radius \citep{sfair12}. That simulation thus predicted a radial color gradient in the ring; however, the model was unable to achieve a steady state, which the authors attribute to limited knowledge of how the particles are removed from the ring, so this expectation should be treated with caution. 
In any case, the physical properties of this unusual moon are of interest for studies of interactions between airless surfaces and the interplanetary environment, the origin and dynamical evolution of circumplanetary dust, and the formation and evolution of small solar system bodies.

The HST observations that discovered Mab could not uniquely constrain its radius and albedo, which are degenerate with respect to the reflected flux of an unresolved body \citep{showalter06}. Those authors proposed two end-member cases: Mab has a $\sim$6 km radius with a Miranda-like icy surface and an albedo of $\sim$0.46, or Mab has a $\sim$12 km radius and an albedo of $\sim$0.1 like its other nearest neighbor, Puck. Previous near-infrared searches with Keck did not detect the moon; the upper limit of \citet{depater06} led to the suggestion that Mab was an icy body like Miranda, with substantial water-ice absorption, and the more sensitive upper limit of \citet{paradis19} required that Mab's radius be $\lesssim$6 km and its surface as dark at 1.6 $\mu$m as Puck \citep[albedo $\lesssim$0.11;][]{paradis23}. Strong near-IR absorption from the crystalline water ice band at 1.65 $\mu$m is responsible for a factor-of-four darkening in the near-IR spectra of Pluto's outer satellites Nix and Hydra \citep{cook18}, and this absorption band could explain such a low H-band flux. Note that the decrease in brightness between visible and 1.6 $\mu$m wavelengths from amorphous water ice absorption cannot account for this factor of four \citep[][]{mastrapa08, terai16}, which would imply that Mab's surface is overlain with crystalline rather than amorphous water ice.

Here we present the first detection of Mab at near-infrared wavelengths. We also report the first near-IR detection of the tiny Portia group satellite Perdita, which to date has been observed only in a reanalysis of Voyager 2 data \citep{karkoschka01b} and the same HST dataset that first detected Mab \citep{showalter06}. Our dataset and data proceessing steps are described in Section \ref{s:observations}. We present photometry of Mab and Perdita in Section \ref{s:results}, and compare the derived infrared reflectance of those two moons with the spectra of Miranda and Puck in Section \ref{s:discussion}.

\section{Observations and Data Processing}
\label{s:observations}

This paper makes use of H-band (1.6 $\mu$m) and Kp-band (2.1 $\mu$m) data taken on 28 October 2019 by the narrow camera of the NIRC2 instrument coupled to the Adaptive Optics (AO) system at Keck Observatory, using Uranus itself for wavefront sensing \citep{wizinowich00, vandam04}. The same data were analyzed by \citet{paradis23}, so we refer the reader to that paper for a detailed description of the observation and data reduction procedures. The standard star was FS2 from the UKIRT Faint Standard Star list \citep{hawarden01},\footnote{\url{https://www2.keck.hawaii.edu/inst/nirc/UKIRTstds.html}} and standard photometric calibration procedures resulted in calibration constants $C_H = 8.87\times10^{-17}$ and $C_{Kp} = 6.46\times10^{-17}$ to convert from 1 count per second to 1 \ergsec{}. We derived the photometric uncertainty by looking at the frame-by-frame surface brightness variations of cloud-free regions of Uranus's disk, which are very stable over the timescales considered here. We determined the surface brightness of the planet in two 20 $\times$ 20 pixel regions of the disk in both H- and Kp-band, one just north of Uranus's equator and one just south of the North Pole. The night was very photometrically stable, and the frame-by-frame variance displayed no clear temporal trends in either H- or Kp-band in either region; the fluctuations looked like Gaussian noise with standard deviations of 
\input{output/urh_photometric_uncertainty_percent.txt}\% in H-band and \input{output/urk_photometric_uncertainty_percent.txt}\% in Kp-band. These numbers agree with the spread in the fluxes derived from the three standard star frames, and we take them to be the 1$\sigma$ photometric calibration uncertainty. The uncertainty in the true flux of the standard star FS2 is less than one part per thousand \citep{hawarden01}.

\begin{figure}
\includegraphics[width=0.5\textwidth]{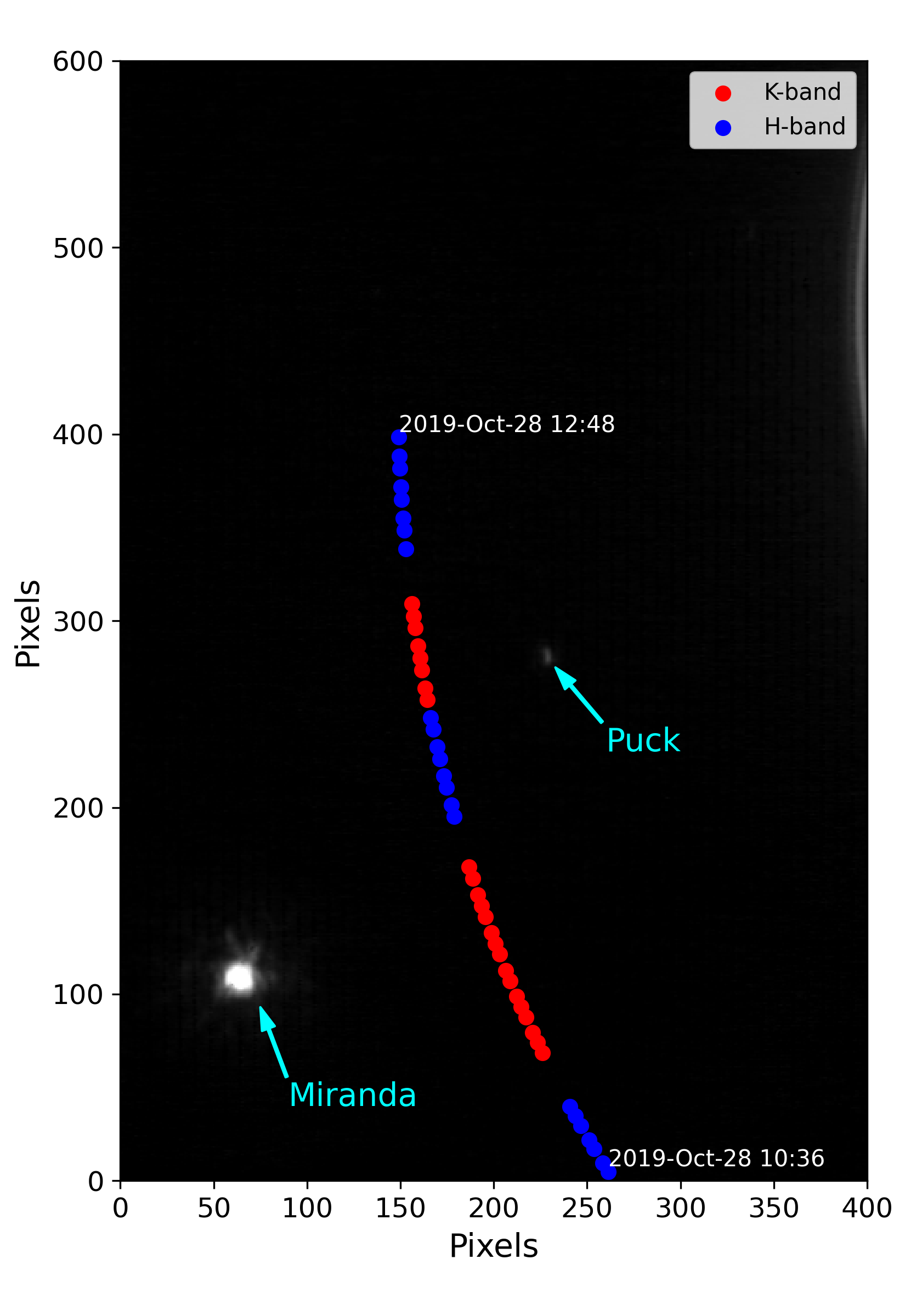}
\includegraphics[width=0.5\textwidth]{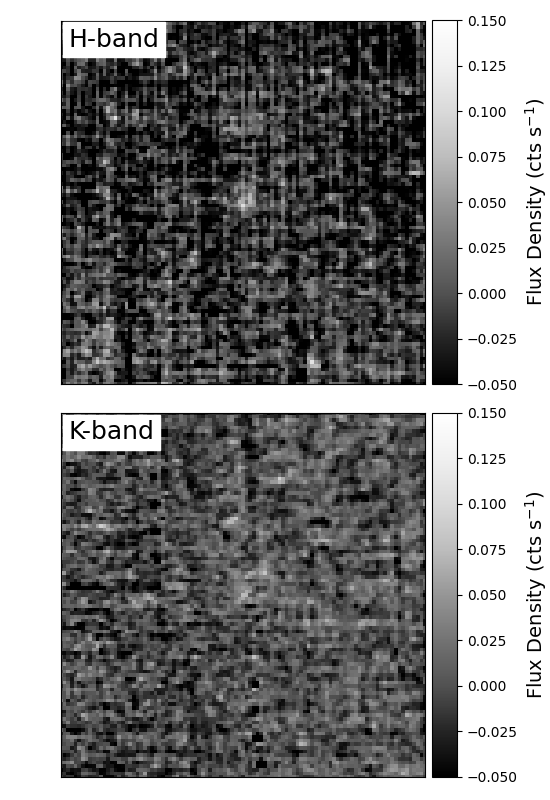}
\caption{\textbf{Left:} A single NIRC2 H-band exposure overlaid with the expected position of Mab in each of our 23 H-band (blue) and 24 Kp-band (red) exposures. The bright and faint point sources visible in the image are Miranda and Puck, respectively. \textbf{Right:} Detection of Mab in H-band (top panel) and non-detection in Kp-band (bottom panel) using the shift-and-stack technique.\label{fig:detection}}
\end{figure}

Our full dataset consisted of 23 exposures of 120 seconds each in H-band and 24 exposures of 120 seconds each in Kp-band; these were all the frames taken on 28 October 2019 in which Mab's expected position was inside the field of view of the narrow camera (Figure \ref{fig:detection}, left panel).
The proper motion of Mab is roughly 30 mas (3 pixels) per minute, which means that each frame had to be shifted relative to the previous one in accordance with Mab's expected vector of motion to improve the signal-to-noise ratio (SNR). This so-called ``shift-and-stack" technique has been used by other authors to improve the SNR of moving point sources, including to measure photometry of other small moons of Uranus with NIRC2 \citep[][]{paradis19, paradis23}. Our shift-and-stack procedure differs in several ways from the one implemented by those authors, and we describe it fully in Supplementary Material \ref{s:shiftandstack}. The application of shift-and-stack resulted in one stacked image per filter at the expected position of Mab, representing effective integration times of 46 minutes in H-band and 48 minutes in Kp-band.

\section{Results}
\label{s:results}

An image of our H-band detection of Mab is shown in the upper right panel of Figure \ref{fig:detection}. The detection has a low SNR, but we are confident that it is real for three reasons. First, a tentative detection of Mab can be made separately from the first 11 frames and the last 11 frames, meaning that an artifact in a single frame or 8-frame group is not producing a false positive. Second, the detection is located in the appropriate position on the sky relative to Miranda and Puck, as discussed below. Third, we performed an experiment to test the false positive rate of our shift-and-stack procedure under the same scattered light and noise conditions as the data. In short, we added random offsets to the expected position of Mab in each frame, then carried out shift-and-stack in the same way as we did for the true expected position. We repeated this procedure with different random offsets 100 times, and found that no noise spikes as bright as our Mab detection appeared anywhere in any of those 100 stacks; see Supplementary Material \ref{s:falsepositives} for more details.

Because Mab was detected at low SNR and within a substantial scattered light background, the flux in the wings of the point-spread function (PSF) could not be measured. We therefore determined the flux of Mab using a flux bootstrapping procedure similar to that applied to Keck-observed point sources by other authors \citep[][]{gibbard05, depater14b, molter19, paradis23}, the specifics of which we summarize in Supplementary Material \ref{s:bootstrapping}. We find the total flux of Mab to be
(\input{output/Mab_urh_flux.txt} $\pm$ \input{output/Mab_urh_fluxerr.txt})$\times10^{-16}$ \ergsec{} 
in H-band, where the error bars indicate a 1-$\sigma$ uncertainty, computed by taking the quadrature sum of the $\sim$13\% flux bootstrapping error and the $\sim$2\% photometric calibration error. 

It is typical to express visible and infrared fluxes of solar system bodies in terms of the unitless quantity I/F, which is defined as
\begin{equation}
\frac{I}{F} = \frac{R^2}{\Omega} \frac{F}{F_\odot}
\end{equation}
where $R$ is the heliocentric distance to the body in AU, $\Omega$ is the solid angle of the body $F$ is the observed flux of the body, and $\pi F_\odot$ is the incident flux of the Sun at the distance to Earth in the same filter and the same units as $F$ \citep{hammel89}. In the case that the shape of the body, and therefore $\Omega$, is unknown, the integrated I/F is used instead, which is defined as the sky-projected area the moon would have if it were a perfect Lambertian reflector \citep[i.e., an I/F of unity;][]{karkoschka01}. The flux level we observe for Mab corresponds to an integrated I/F of 
\input{output/Mab_urh_intif.txt} $\pm$ \input{output/Mab_urh_intiferr.txt} km$^{2}$ 
at 1.6 $\mu$m. Note that the Sun-target-observer phase angle $\alpha$ of these observations was only 0.02$^\circ$, so the opposition surge was likely substantial. The sub-observer longitude of Mab passed through 270$^\circ$ during the $\sim$2 hours of observations. Therefore, assuming Mab is tidally locked, we observed mostly its trailing hemisphere and north pole.

\begin{figure}
\includegraphics[width=0.5\textwidth]{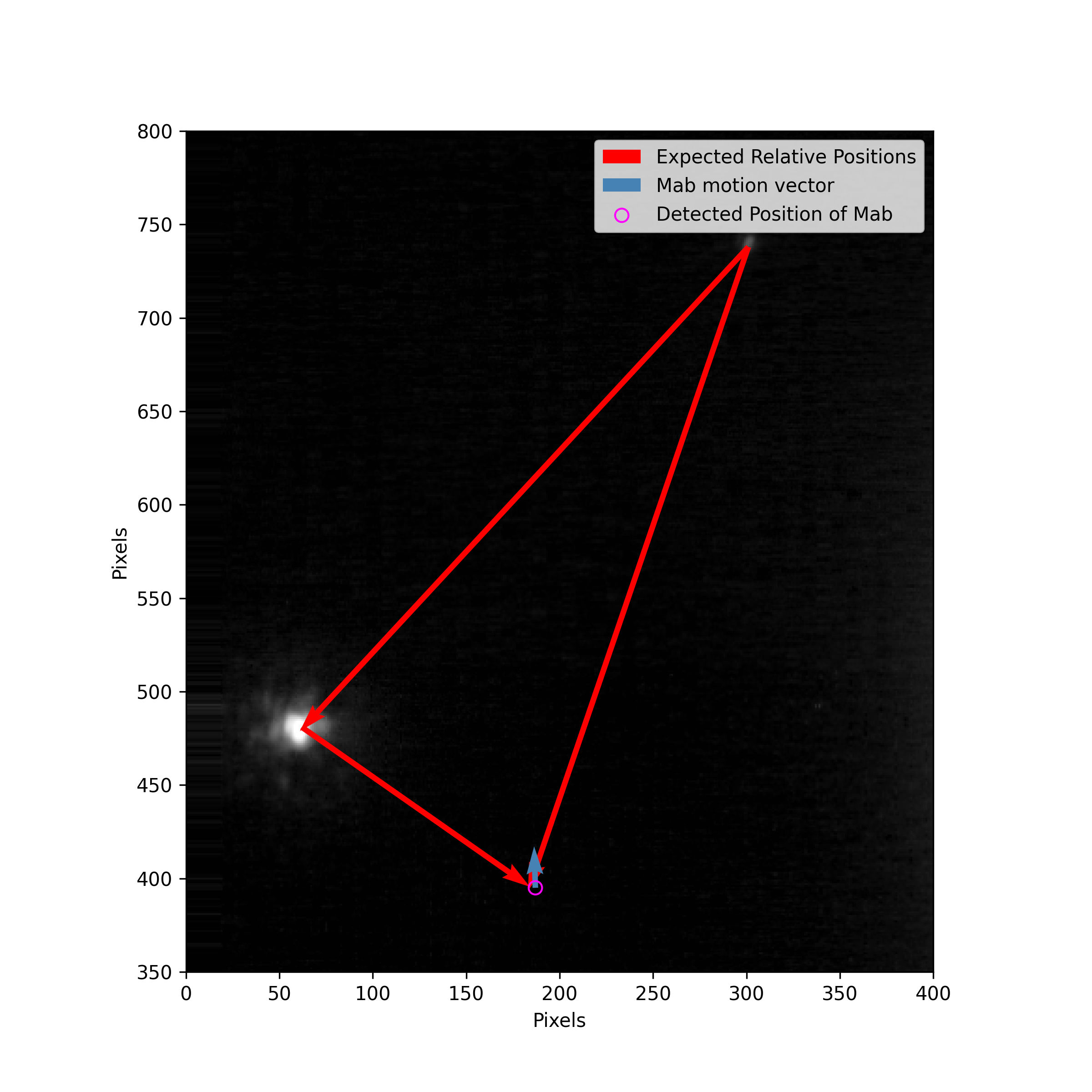}
\includegraphics[width=0.5\textwidth]{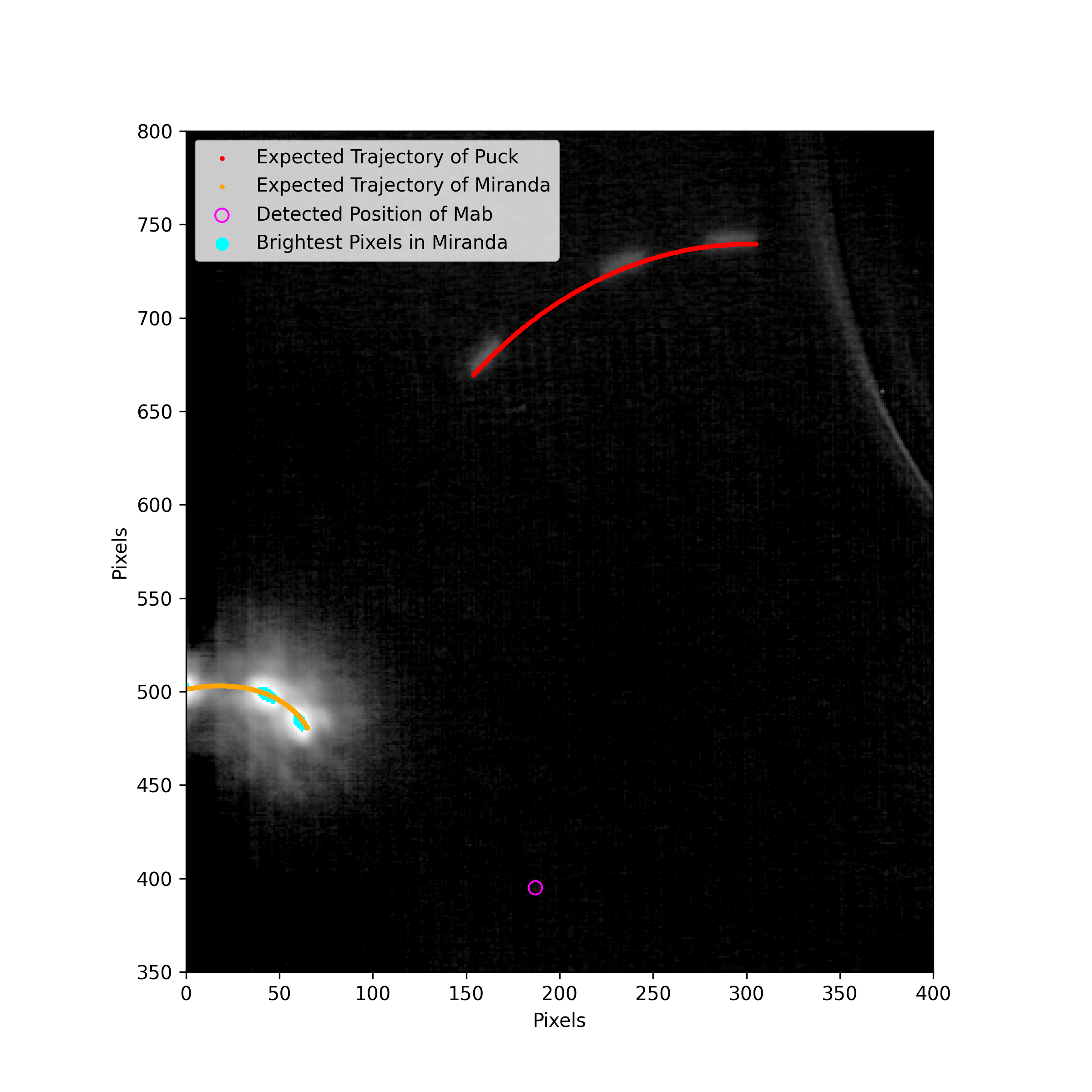}
\caption{Two visualizations of Mab's expected position. \textbf{Left:} A single NIRC2 H-band exposure overlaid with the expected relative positions of Mab, Miranda, and Puck (red arrows) as well as the detected position of Mab (purple circle). The expected vector of motion of Mab is also overlain (blue arrow). \textbf{Right:} The same shifted-and-stacked detection image of Mab shown in the right panel of Figure \ref{fig:detection}, but zoomed out and on a different color scale to show the apparent trajectories of Puck and Miranda. Mab's position is marked with a purple circle. The expected motion of Puck and Miranda relative to Mab are plotted as red and orange lines, respectively. Miranda is still overexposed in this image stretch, so the brightest pixels in Miranda's track are highlighted in cyan.}
\label{fig:astrometry}
\script{relative_position_miranda_puck.py}
\end{figure}

The observed position of Mab is the same as expected based on the orbital parameters in the JPL Horizons system\footnote{\url{https://ssd.jpl.nasa.gov/horizons/app.html\#/}} to within one Keck resolution element ($\pm$20 mas or $\pm$270 km). We determined this based on the relative positions of Mab, Miranda, and Puck, the latter two of which can be seen easily in a single 120-second frame; see Figure \ref{fig:astrometry}. The advantage of using the positions of other moons as a reference for astrometry is that it does not depend on a navigation solution for the planet's disk nor the accuracy of the telescope pointing. Previously-reported large residuals in Mab's orbit solution \citep{showalter06} were caused by a 0.13\% plate scale error in the HST HRC/Clear filter \citep{showalter19}, and this has already been corrected in the JPL Horizons system. Our detection confirms the accuracy of that solution to within the above-stated uncertainty.

A very faint region of positive flux was observed at the expected position of Mab after stacking the Kp-band data (see Figure \ref{fig:detection}). This feature is not clearly distinguishable by eye from positive noise spikes elsewhere in the frame. The false positive rate experiment discussed in Supplementary Material \ref{s:falsepositives} shows that a noise spike of similar brightness would appear within $\sim$50 pixels of the expected position $\sim$\input{output/perturbation_percent_higher_K.txt}\% of the time. We therefore consider this a non-detection. Based on that experiment, we can see that a point source whose PSF had a maximum brightness value of 5 times the per-pixel RMS noise would have been clearly detectable. By this definition, the 5-$\sigma$ upper limit to Mab's Kp-band flux is
$\lesssim$\input{output/Mab_urk_flux.txt}$\times10^{-16}$ \ergsec{}. 
This corresponds to an integrated I/F of 
$\lesssim$\input{output/Mab_urk_intif.txt} km$^{2}$ 
at 2.1 $\mu$m, in line with the previous K-band 3$\sigma$ upper limit of $\lesssim$30 km$^2$ derived from single 120-second exposures \citep{depater06}.

We also report the first infrared detection of Perdita in H-band with a flux of
(\input{output/Perdita_urh_flux.txt} $\pm$ \input{output/Perdita_urh_fluxerr.txt})$\times10^{-16}$ \ergsec{}.
This translates to an integrated I/F of \input{output/Perdita_urh_intif.txt} $\pm$ \input{output/Perdita_urh_intiferr.txt} km$^2$ at $\alpha=0.02^\circ$. We derive a Kp-band upper limit of $\lesssim$\input{output/Perdita_urk_flux.txt}$\times10^{-16}$ \ergsec{}. The sub-observer longitude was near 200$^\circ$ at the time of observation, i.e., near the anti-Uranus point.
Like Mab, Perdita was detected at its expected orbital longitude to within one Keck H-band resolution element, or ($\pm$20 mas or $\pm$270 km).

\section{Discussion}
\label{s:discussion}

\begin{figure}
\includegraphics[width=0.7\textwidth]{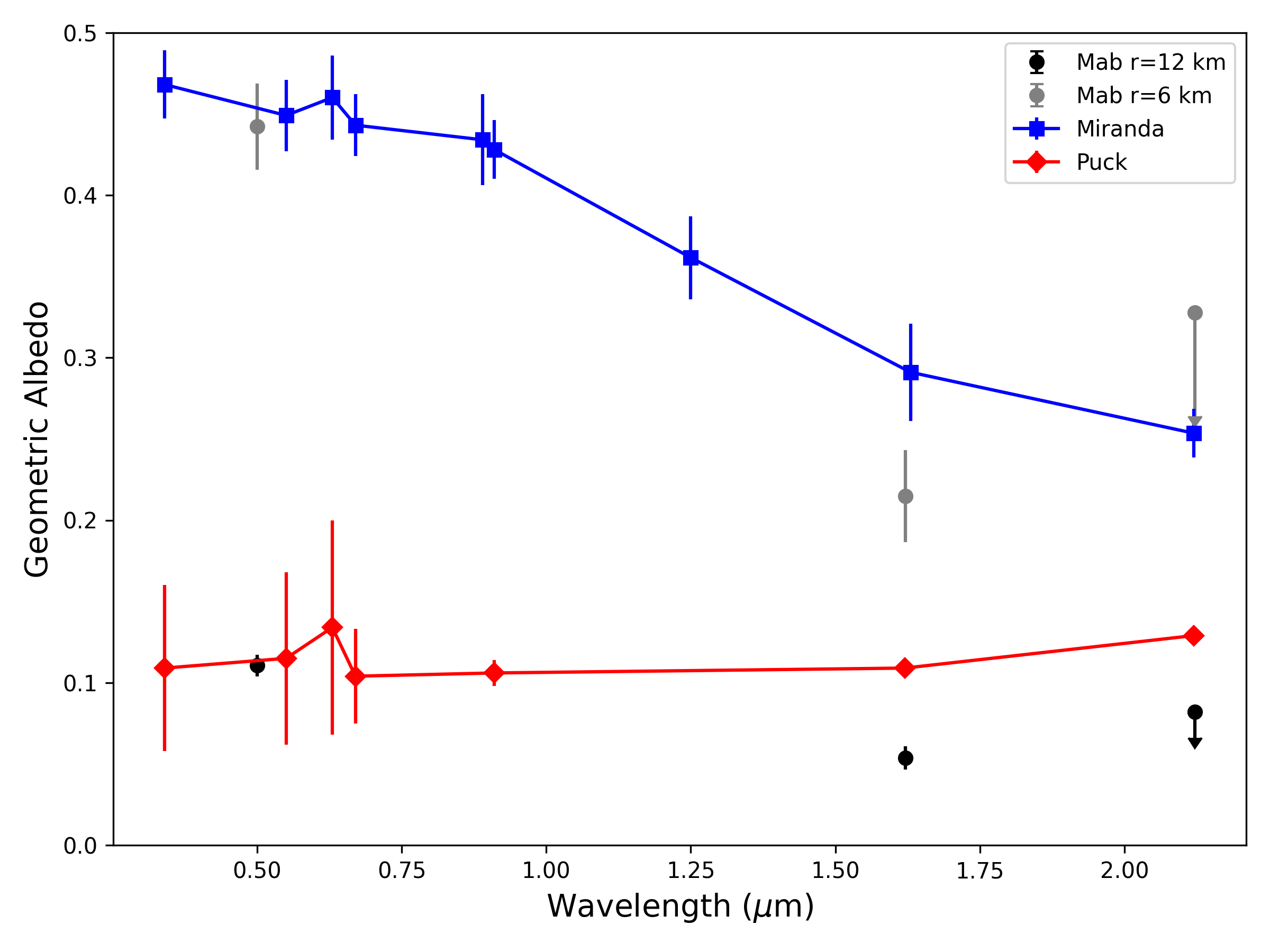}
\caption{Comparison between the spectra of Mab, Miranda, and Puck. HST data for Puck and Miranda are from \citet{karkoschka01} ($\alpha$ = 0.034$^\circ$). Keck H- and Kp-band data for Miranda are from \citet{gibbard05} (rescaled from $\alpha$ = 1.9$^\circ$ to Mab's observed $\alpha$ = 0.02$^\circ$ using the \citet{karkoschka01} photometric model). Keck H- and Kp-band data for Puck are from \citet{paradis23} ($\alpha$ = 0.02$^\circ$). HST Mab data are from \citet{showalter06} ($\alpha$ = 0.08$^\circ$).}
\label{fig:spectrum}
\script{compute_if.py}
\end{figure}

The infrared brightness of Mab is approximately a factor of 2 fainter than observed by \citet{showalter06} with HST in the HRC/Clear filter centered at 0.5 $\mu$m.\footnote{A factor of 1.6 discrepancy between the photometric calibration of HST's HRC/clear filter and the other filters was seen by \citet{showalter06}; those authors multiplied all of their results by this factor of 1.6, and we are using the published final (i.e., corrected) values here.} We compare the I/F of Mab in the three available filters (HST HRC/clear, H, Kp) with I/F values in the same filters for Puck and Miranda \citep{karkoschka01, gibbard05, paradis23} in Figure \ref{fig:spectrum} assuming Mab has either a 6-km or 12-km radius. Ratios between these filters are compiled in Table \ref{tab:color}; large ratios likely indicate an icy surface, as explained in the Introduction. 
The visible/H ratio of Mab is 
closer to Miranda's visible/H ratio than that of Puck. We therefore prefer the interpretation that Mab is a 6-km body with a Miranda-like water-ice-rich surface; however, the alternative interpretation that Mab is a 12-km body with a Puck-like surface cannot be definitively ruled out by the available data.

\input{output/color_table.tex}

\citet{karkoschka01b} determined a radius of 15 $\pm$ 3 km for Perdita assuming its reflectivity at $\alpha=16^\circ$ was 0.033, the average value for the Portia group. For a more direct comparison to our H-band observations, we rescaled Perdita's visible reflectivity from $\alpha = 16^\circ$ to $\alpha = 0.02^\circ$ according to the average Portia group phase curve given by \citet{karkoschka01}. This yields an integrated I/F of $54 \pm 22$ km$^2$ at visible wavelengths, and this is the value we used to derive the color information in Table \ref{tab:color}. The large uncertainty on Perdita's visible/infrared color precludes any definitive statements about that moon's composition.

Our reported H-band flux of 
(\input{output/Mab_urh_flux.txt} $\pm$ \input{output/Mab_urh_fluxerr.txt})$\times10^{-16}$ \ergsec{} 
for Mab is seemingly in conflict with the 3$\sigma$ upper limit of $2.4 \times 10^{-16}$ \ergsec{} derived previously \citep[][]{paradis19}, which we have rescaled to account for updated photometric calibration constants reported in \citet{paradis23}. However, those observations were taken at a phase angle of $2.02^\circ$. If we rescale the limit to the $0.02^\circ$ phase angle of our observations according to the \citet{karkoschka01} photometric model for Miranda, the limit would be $3.9 \times 10^{-16}$ \ergsec{}, which is 2.8$\sigma$ from our value. One possible explanation for this factor-of-1.6 discrepancy is that Mab has an oblong shape, and the orbital longitude of our observations was more conducive to observing a larger cross-sectional area than that of \citet{paradis19}. The similarly-sized Saturnian moons Calypso and Atlas, for example, have triaxial ellipsoid radii $a:b:c$ of approximately 5:3:2 and 2:2:1 \citep{thomas20}, so it is reasonable to expect factor-of-two differences in cross-sectional area. However, more detailed studies of the light curve of Mab over its orbit would be required to achieve trustworthy constraints on the moon's shape.

Although the present results are suggestive of compositional differences among Uranus's small moons, the large uncertainty on the radii of both Mab and Perdita limit the physical interpretation available from photometry alone. Studies in the near future with the James Webb Space Telescope (JWST) will remedy this: JWST's mid-infrared instrument is sensitive enough to likely detect thermal emission from both moons \citep{wright23}, permitting size determinations via thermal radiometry, as has been routinely performed on asteroids and Kuiper Belt Objects \citep[see][for definition and examples of this technique]{muller02, stansberry08, vilenius18}.

%The \citet{paradis23} requirement that Mab's H-band geometric albedo be less than 0.11 for a 6-km effective radius is disfavored by our data at the 3.5$\sigma$ level.

%The photometric model of \citet{karkoschka01} defines the I/F of a moon as the product of four normalized functions and a constant:
%\begin{equation}
%	\label{eq:if}
%	I/F(l, \alpha, \lambda) = f_1(l) f_2(\alpha) f_3(\lambda) f_4(\lambda) (I/F)_0
%\end{equation}
%where $f_1(l)$ is the lightcurve as a function of orbital longitude $l$ with an orbital average of unity, $f_2(\alpha)$ is the phase function scaled to unity at phase angle $\alpha = 0$, $f_3(\lambda)$ is the solar spectrum (the ``continuum'') at the observed wavelength $\lambda$ scaled to unity at 550 nm, $f_4(\lambda)$ is the band/continuum ratio, used to account for wavelength-dependent absorption features in the spectrum of the moon, and $(I/F)_0$ is the reflectivity at $\alpha = 0$ and $\lambda = 550$ nm, averaged over orbital longitude.

\vspace{1cm}

This research was supported by NASA grant NNX16AK14G through
the Solar System Observations (SSO) program to the University of
California, Berkeley.

The data presented herein were obtained at the W. M. Keck Observatory, which is operated as a scientific partnership among the California Institute of Technology, the University of California and the National Aeronautics and Space Administration. The Observatory was made possible by the generous financial support of the W. M. Keck Foundation.

The authors wish to recognize and acknowledge the very significant cultural role and reverence that the summit of Maunakea has always had within the indigenous Hawaiian community.  We are most fortunate to have the opportunity to conduct observations from this mountain.

We thank the two referees, Luke Dones and one anonymous person, for their insightful comments and suggestions, which substantially improved the manuscript.

This study was carried out using the reproducibility software
\href{https://github.com/showyourwork/showyourwork}{\showyourwork}
\citep{luger21}, which leverages continuous integration to
programmatically download the data from
\href{https://zenodo.org/}{zenodo.org}, create the figures, and
compile the manuscript. Each figure caption contains two links: one
to the dataset stored on zenodo used in the corresponding figure,
and the other to the script used to make the figure (at the commit
corresponding to the current build of the manuscript). The git
repository associated to this study is publicly available at
\url{https://github.com/emolter/mab}, and the release
v0.0.2 allows anyone to re-build the entire manuscript. The datasets
are stored at \url{https://doi.org/10.5281/zenodo.7689583}.

\software{
Astropy \citep{astropy:2013, astropy:2018, astropy:2022},
Astroquery \citep{ginsburg19},
matplotlib \citep{matplotlib07},
numpy \citep{numpy20},
photutils \citep{photutils140},
pylanetary \citep{pylanetary23},
scikit-image \citep{skimage14},
scipy \citep{scipy20},
shift\_stack\_moons \citep{shift_stack_moons23}
}

\facility{Keck:II (NIRC2)}

\appendix

\restartappendixnumbering

\section{Shift-and-Stack Procedure}
\label{s:shiftandstack}

To improve the SNR on Mab, individual 120-second frames were shifted according to the expected position of the moon relative to Uranus and then co-added; details of this procedure are presented here. The accuracy of the shift-and-stack procedure depends on the rotation angle of the detector, so we corrected for the 0.26$^\circ$ angular offset between NIRC2's sky north position and true sky north \citep{service16}. First, we aligned the position of Uranus in all the frames using the \texttt{chi2\_shift} function of the \texttt{Astropy}-affiliated \texttt{image\_registration} Python package\footnote{\url{https://github.com/keflavich/image_registration}}, which implements the single-step discrete Fourier transform (DFT) algorithm for efficient sub-pixel image registration \citep{guizarsicairos08}. In Kp-band, cloud features and moons are brighter than the planet's disk, so we first applied a Canny edge-detection filter using the \texttt{skimage} Python package \citep{skimage14} to enhance the rings of Uranus, then ran \texttt{chi2\_shift} on those edges. This alignment method was found to produce very sharp stacked images of the rings, so we were confident that it aligned the frames accurately. The implementation of this alignment step permitted us to co-add all three H-band observing blocks (and two Kp-band observing blocks) shown in Figure \ref{fig:detection}, between which Uranus was repositioned on the detector by tens of pixels.  Second, we found the median of these planet-aligned frames and subtracted it from each frame; this step suppresses the signal of non-moving sources (in this case, the planet and rings), decreasing the effects of scattered light. Third, the expected offset between Mab and Uranus at the time\footnote{Start time plus half the exposure time} of each exposure was taken from JPL Horizons, accessed using the \texttt{Astroquery} Python package \citep{ginsburg19}. We converted the expected x,y offset from units of arcseconds to pixels assuming the pixel scale of the NIRC2 narrow camera to be 0.009971\arcsec \citep{service16}. Finally, we applied the shifts using a 2-D fast Fourier transform (FFT) sub-pixel image shift, again implemented with \texttt{image\_registration}, and summed all the frames to produce a single stacked frame. Our shift-and-stack code is publicly available on GitHub \citep{shift_stack_moons23}.

\section{Flux Bootstrapping Technique}
\label{s:bootstrapping}

Using the \texttt{astropy}-affiliated \texttt{photutils} package \citep{photutils140}, we measured the total flux inside many circular apertures of different radii $r_i$ centered on Mab; let us call Mab's flux for the $i$th radius $F_{M i}$. The flux of the background must be subtracted in order to accurately determine $F_{M i}$; we defined the background region using a circular annulus centered on Mab with inner radius $r_i + 5$ pixels and outer radius $r_o$. The background flux level $F_{B o}$ was taken to be the median of pixels within the background annulus multiplied by the area of the Mab aperture $\pi r_i^2$. We also varied $r_o$ over a range of values to test the sensitivity of our flux estimate to the choice of background region. We then measured the flux inside apertures of the same size but centered on the standard star; call these $F_{\star i}$. The same background regions were also applied to the standard star, yielding one star flux measurement $F_{\star i,o}$ for each inner and outer radius pair. Comparing $F_{\star i,o}$ with the true flux of the standard star $F_\star$ (as measured in a very large aperture of radius $\gtrsim$100 pixels) yielded a correction at given values of $r_i$ and $r_o$ given by simply $C_{psf,i,o} = F_{\star i,o}/F_\star$.  The flux of Mab $F_{i,o}$ for a given $r_i, r_o$ pair is given by $F_M = (F_{M i} - F_{B o}) / C_{psf,i,o}$.
The final flux estimate of Mab is given by the mean of $F_{i,o}$ over a reasonable range of $r_i$ and $r_o$ values, which we take to be 3-10 pixels and 15-50 pixels, respectively. The uncertainty is given by the standard deviation of $F_{i,o}$, which was $\approx$15\% in both H- and Kp-band. This error was added in quadrature with the photometric uncertainty.

%\begin{figure}
%\includegraphics[width=0.3\textwidth]{figures/example_aperture.png}
%\includegraphics[width=0.7\textwidth]{figures/photometry_vs_region.png}
%\caption{\textbf{Left:} Example aperture (solid blue line) and background annulus (dot-dashed red line) centered on Mab. The values of $r_i$ and $r_o$ in this example are 7 and 45 pixels, respectively. \textbf{Right:} Derived flux of Mab as a function of $r_i$ and $r_o$.\label{fig:wing}}
%\end{figure}

\section{False Positive Rate Experiment}
\label{s:falsepositives}

\begin{figure}
\includegraphics[width=1.0\textwidth]{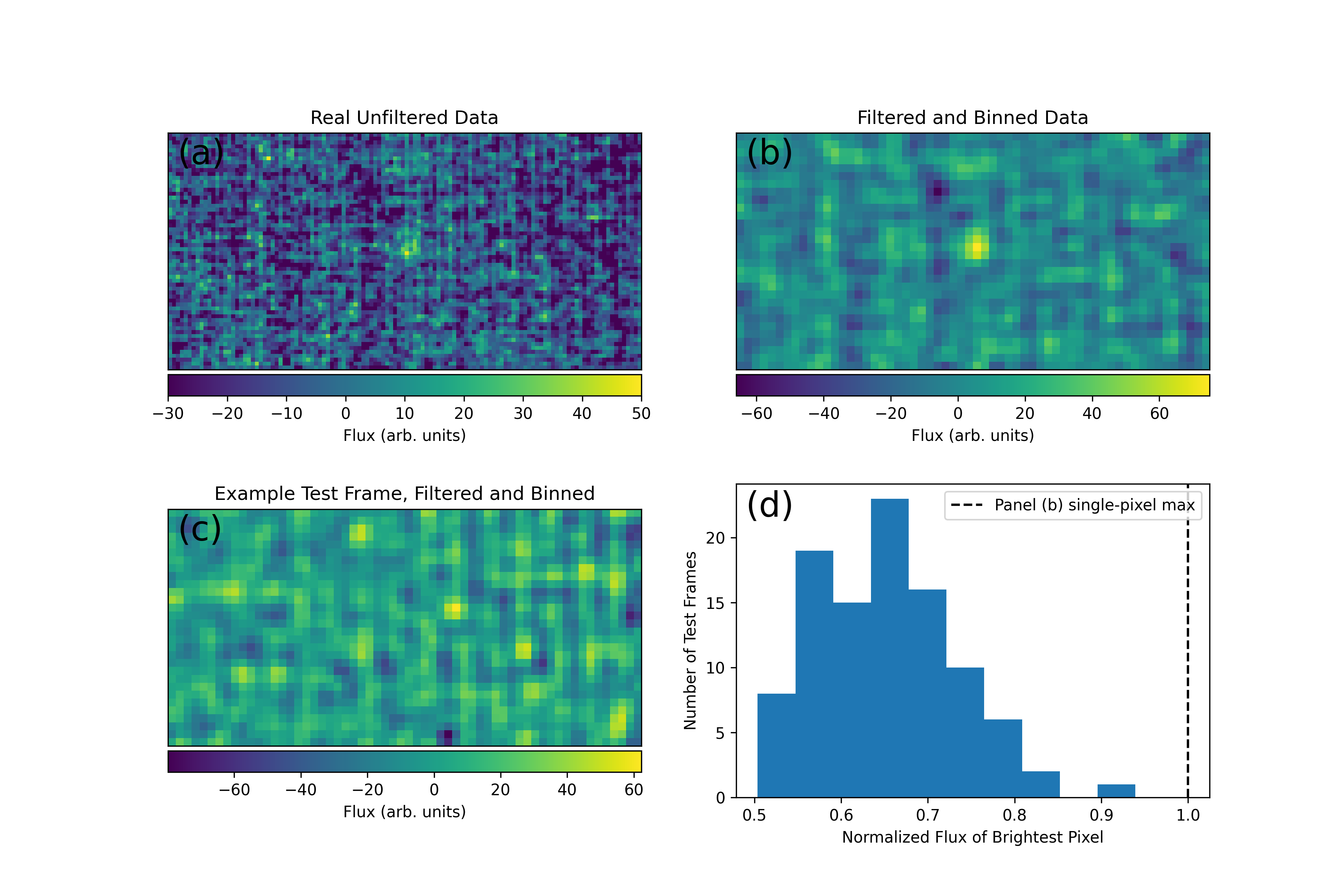}
\caption{Results of the false positive rate experiment for H band. \textbf{(a)} Detection of Mab shown in Figure \ref{fig:detection}. \textbf{(b)} Panel (a) after filtering and binning were applied. \textbf{(c)} Example filtered and binned randomized stack. \textbf{(d)} Histogram of the SNR of the brightest pixel in the filtered and binned test frames; the real data are demarcated with a black dashed line.}
\label{fig:randomstack}
\script{plot_perturbation_experiment.py}
\end{figure}

\begin{figure}
\includegraphics[width=1.0\textwidth]{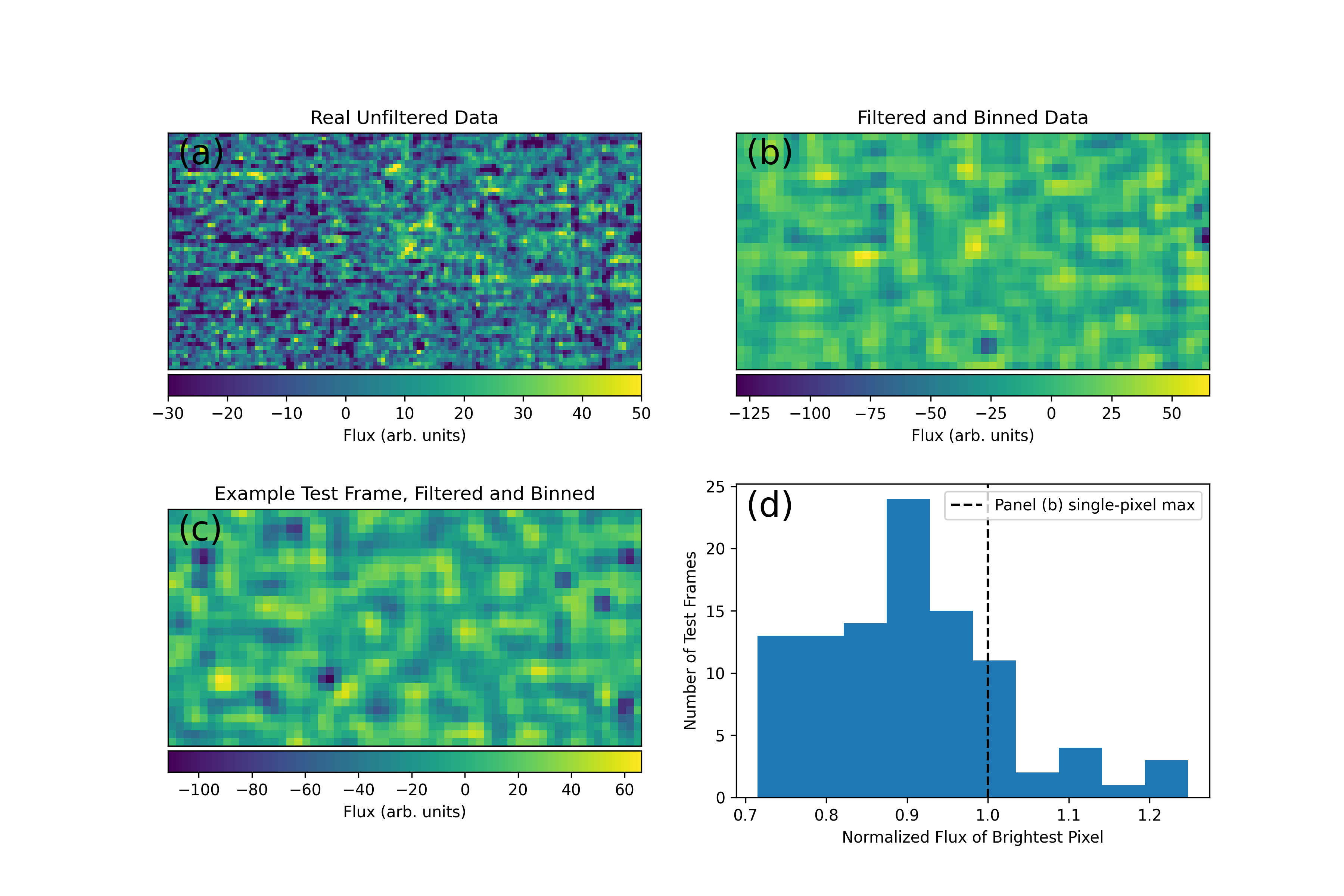}
\caption{Results of the false positive rate experiment for Kp band. See caption of Figure \ref{fig:randomstack} for a description. }
\label{fig:randomstack_K}
\script{plot_perturbation_experiment.py}
\end{figure}

Mab was observed in the lower-left quadrant of the NIRC2 narrow camera, which is known to have non-Gaussianity in its noise statistics,\footnote{https://www2.keck.hawaii.edu/inst/nirc2/ObserversManual.html} including vertical stripes of correlated noise. Given the low SNR of the detection, it is natural to wonder whether the shift-and-stack technique is injecting false positives by stacking noise spikes atop one another. To determine whether this is the case, we performed an experiment to test the false positive rate within the same detector noise and scattered light environment of the data, as follows. The expected position of Mab was shifted by a random x,y offset of mean amplitude 20 pixels in each frame, then the shift-and-stack algorithm was re-run assuming these new randomly-selected shifts, and the output frame was saved. This experiment was repeated 100 times with different random offsets applied. We then searched for point sources in these 100 test stacks by applying a center-surround (difference-of-Gaussians) filter with an FWHM of 4 pixels for the inner Gaussian (this corresponds roughly to the diffraction-limited beam size of Keck in H-band) and an FWHM of 12 pixels for the outer Gaussian. The center-surround filter is the appropriate choice because it accentuates features at the desired spatial frequency and downweights low spatial frequency features. The visual systems of humans and other animals employ center-surround antagonism \citep{graham06}, so this technique is a reasonable proxy for ``by-eye'' detection. We binned the filtered images by a factor of 2 in each direction so that a single pixel was more closely representative of a feature at the desired spatial frequency. Finally, we computed the amplitude of the brightest pixel in each binned frame. Identical filtering, binning, and SNR calculations were run on the real data. The detection of Mab in the real data was found to be brighter than any peak in any of the 100 test frames in H-band (see Figure \ref{fig:randomstack}). In Kp-band, we find that a noise spike with an amplitude higher than our tentative detection is present in \input{output/perturbation_percent_higher_K.txt} out of 100 frames (Figure \ref{fig:randomstack_K}), so we label it a non-detection. We tested other choices of inner and outer FWHM for the center-surround filter and different amounts of binning, and found that the results remained similar within reasonable ranges of these parameters. 

\bibliography{references}

\end{document}

%% file: output/Mab_urh_intif.txt
24

%% file: output/Mab_urh_intiferr.txt
3

%% file: output/Mab_urk_intif.txt
37

%% file: output/Perdita_urh_intif.txt
31

%% file: output/Perdita_urh_intiferr.txt
3

%% file: output/urh_photometric_uncertainty_percent.txt
2.1

%% file: output/urk_photometric_uncertainty_percent.txt
5.0

%% file: output/Mab_urh_flux.txt
6.1

%% file: output/Mab_urh_fluxerr.txt
0.8

%% file: output/perturbation_percent_higher_K.txt
16

%% file: output/Mab_urk_flux.txt
3.6

%% file: output/Perdita_urh_flux.txt
7.8

%% file: output/Perdita_urh_fluxerr.txt
1.0

%% file: output/Perdita_urk_flux.txt
3.8

%% file: output/color_table.tex
\begin{table}
\caption{Available color information on Mab and Perdita compared with Puck and Miranda. See Figure \ref{fig:spectrum} caption for data sources. \label{tab:color}}
\begin{tabular}{ccc}
\hline \hline
Moon & 0.5 $\mu$m / 1.6 $\mu$m & 1$\sigma$ error \\
\hline
Miranda & 1.61 & 0.18 \\
Puck & 1.0 & 0.47 \\
Perdita & 1.74 & 0.74 \\
Mab & 2.06 & 0.3 \\
\hline
\end{tabular}
\end{table}